\documentclass[prb,aps,onecolumn,amsmath]{revtex4}
\usepackage{graphicx}
\usepackage{dcolumn}
\usepackage{bm}

\begin{document}
\title{Current-controlled magnetization dynamics in the spin-flip transistor}

\author{Xuhui Wang}
\author{Gerrit E. W. Bauer}%
\affiliation{Kavli Institute of NanoScience, 
Delft University of Technology, 2628 CJ Delft, The Netherlands}%
\author{Teruo Ono}%
\affiliation{Institute for Chemical Research, Kyoto University, Uji Kyoto 611-0011, Japan}%
\date{\today}

\begin{abstract}
The current driven magnetization dynamics of a thin-film,
three magnetic terminal device (spin-flip transistor) is
investigated theoretically. We consider a magnetization
configuration in which all magnetizations are in the device plane,
with source-drain magnetizations chosen fixed and antiparallel,
whereas the third contact magnetization is allowed to move in a weak
anisotropy field that guarantees thermal stability of the
equilibrium structure at room temperature. We analyze the
magnetization dynamics of the free layer under a dc source-drain
bias current within the macrospin model and magneto-electronic
circuit theory. A new tunable two-state behavior of the
magnetization is found and the advantages of this phenomenon and
potential applications are discussed.
\end{abstract}

\maketitle

\section{Introduction}
\begin{figure}
\begin{center}
\includegraphics[scale=0.3]{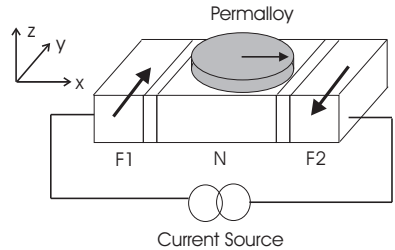}
\end{center}
\caption{The model system contains a normal metal sandwiched by two
ferromagnetic leads and a circular soft ferromagnet film (e.g.,
permalloy) on top of the normal metal. The magnetization unit
vectors $\textbf{m}_{1}$, $\textbf{m}_{2}$, and $\textbf{m}_{3}$ are
initially aligned in the same, i.e., $x-y$ plane.}
\label{fig:geometry}
\end{figure}

The current induced magnetization excitation predicted by Slonczewski
and Berger\cite{slonczewski,berger} has attracted considerable attention
and the prediction of current-induced magnetization reversal has been confirmed
by many experiments in nano-pillar structure consisting of two ferromagnetic layers
with a high (``fixed") and a low (``free") coercivity, separated by a normal metal
spacer.\cite{katine2000prl,myers2002prl,kiselevnature,rippardprl} Meanwhile, the
investigations of charge and spin transport in thin-film metallic conductors structured
on a planar substrate have also been carried out.
\cite{jedema2001nature,jedema2002nature,vanwees,riken,tinkham,argonne}
The advantages of the planar structures are the flexible design and the relative
ease to fabricate multi-terminal structures.\cite{bauer-spintorque}
Recently, non-local magnetization switching in a lateral spin valve structure has been
demonstrated.\cite{kimura-condmat2005} In the present article we present a
 theoretical study on the dynamics of a lateral spin valve consisting of a normal
metal film that is contacted by two magnetically hard ferromagnets.
As sketched in Fig.~\ref{fig:geometry}, a slightly elliptic and
magnetically soft ferromagnetic film is assumed deposited on top of
the normal metal to form a spin-flip transistor.\cite{brataasprl}
The magnetization direction of the source-drain contacts lies
antiparallel to each other in the plane of the magnetization of the
third (free) layer. The configuration in which the source-drain
contact magnetizations are oriented perpendicular to the plane is
considered elsewhere.\cite{xwang} A convenient and accurate tool to
study the dynamic properties of our device is the magnetoelectronic
circuit theory (MECT) for charge and spin transport\cite{brataasprl}
combined with the Landau-Lifshitz-Gilbert equation in the macrospin
model. The spin flip scattering in normal and ferromagnetic metals
and the spin-pumping effect are also taken into
account.\cite{yaroslavprl,yaroslavrmp}

The article is organized as the follows: In \S
\ref{sec:formalism}, we briefly review the MECT and
Landau-Lifshitz-Gilbert equation for the macrospin model. In \S
\ref{sec:stt}, calculations of the spin transfer torque for our
device are presented. \S \ref{sec:thermal} is devoted to the
discussion of thermal (in)stability and in \S \ref{sec:dyna}
the magnetization dynamics is treated. The conclusions are
summarized in \S \ref{sec:concl}.

\section{\label{sec:formalism} Magneto-Electronic Circuit Theory}

We first consider a ferromagnet-normal metal ($\text{F}|\text{N}$)
interface at quasi-equilibrium. The ferromagnet at a chemical
potential $\mu_{0}^{F}$ and spin accumulation
$\mu_{s}^{F}\textbf{m}$ aligned along the magnetization direction.
The chemical potential and spin accumulation in the normal metal are
denoted by $\mu_{0}^{N}$ and vector $\mathbf{S}$. The charge current
$I_{c}$ (in the unit of Ampere) and the spin current
$\mathbf{I}_{s}$ (in the unit of Joule) entering the normal metal
node are given by,~\cite{brataasprl}
\begin{align}
I_{c}=&\frac{e}{2h}\left[2g(\mu_{0}^{F}-\mu_{0}^{N})+pg\mu_{s}^{F}-pg~(\mathbf{S}\cdot\mathbf{m})\right]\nonumber\\
\mathbf{I}_{s}=&\frac{g}{8\pi}[2p(\mu_{0}^{F}-\mu_{0}^{N})+\mu_{s}^{F}-(1-\eta_{r})(\mathbf{S}\cdot\mathbf{m})]\textbf{m}\nonumber\\
&-\frac{\eta_{r}g}{8\pi}\mathbf{S}-\frac{\eta_{i}g}{8\pi}(\mathbf{S}\times\mathbf{m})~.
\label{eq:current}
\end{align}
From eq. (\ref{eq:current}) we may project out the component of
$\textbf{I}_{s}$ that is perpendicular to the magnetization
direction and governs the spin transfer
torque~\cite{slonczewski,brataasprl}
\begin{equation}
-\mathbf{m}\times\mathbf{I}_{s}\times\mathbf{m}=\frac{\eta_{r}g}{8\pi}[\mathbf{S}-(\mathbf{S}\cdot\textbf{m})\textbf{m}]
+\frac{\eta_{i}g}{8\pi}(\textbf{S}\times\textbf{m})~.
\label{eq:torquegeneral}
\end{equation}
In the above notations, the dimensionless total conductance
$g=g^{\uparrow}+g^{\downarrow}$ and the mixing conductance
are given by Landauer-B\"uttiker formulae, i.e.,
\begin{align}
g^{\uparrow(\downarrow)}&=M-\sum\limits_{nm}|r_{\uparrow(\downarrow)}^{nm}|^{2}~,\nonumber\\
g^{\uparrow\downarrow}&=M-\sum\limits_{nm}r_{\uparrow}^{nm}(r_{\downarrow}^{nm})^{\ast}~.
\end{align}
where $r_{\uparrow(\downarrow)}^{nm}$ is the probability of a spin
up(down) electron in mode $m$ reflected into mode $n$ in the normal
metal and $M$ is the total number of channels. The contact polarization 
is defined by $p=(g^{\uparrow}-g^{\downarrow})/(g^{\uparrow}+g^{\downarrow})$.
The pumping current
generated by the motion of the magnetization is \cite{yaroslavprl}
\begin{equation}
\mathbf{I}_{s}^{(p)}=\frac{\hbar g}{8\pi}\left(\eta_{r}\mathbf{m}\times\frac{d\mathbf{m}}{dt}
+\eta_{i}\frac{d\mathbf{m}}{dt}\right)~.
\end{equation}
We consider the situation in which the dimension of
the normal metal is smaller than the spin-flip length,
so that the spin accumulation does not vary spatially in the node.
\begin{equation}
\mathbf{I}_{s}^{(f)}=\frac{g_{f}}{4\pi}\mathbf{S}
\end{equation}
where $g_{f}=h\nu_{DOS}\text{V}_{N}/\tau_{f}^{N}$, $\nu_{DOS}$ and
$\text{V}_{N}$ are the density of states of the electrons and the
volume of the normal metal, $\tau_{f}^{N}$ is the spin-flip
relaxation time in the normal metal node. The charge and spin
currents entering the normal metal obey the conservation laws
\begin{align}
\sum_{i}I_{c,i}&=0~,\nonumber\\
\sum_{i}\left(\mathbf{I}_{s,i}+\mathbf{I}_{s,i}^{(p)}\right)&=\mathbf{I}_{s}^{(f)}~.
\label{eq:conserv}
\end{align}

\section{\label{sec:stt}Spin Transfer Torque}
In the structure depicted in Fig. \ref{fig:geometry}, the
source-drain magnetizations are aligned anti-parallel along the
$y$-axis in order to inject a large spin accumulation into
$\text{N}$, i.e., $\mathbf{m}_{1}=(0,+1,0)$ and
$\mathbf{m}_{2}=(0,-1,0)$. Connecting the ferromagnets to reservoirs
and applying a bias current $I_{0}$ via the two ferromagnetic leads,
the conservation of charge current dictates that $I_{c,1}=I_{0}$ and
$I_{c,2}=-I_{0}$ at the $\text{F}1|\text{N}$ and
$\text{F}2|\text{N}$ interfaces, which gives
\begin{align}
\mu_{0}^{F1}-\mu_{0}^{N}=-(\mu_{0}^{F2}-\mu_{0}^{N})=\frac{I_{0}h}{ge}+\frac{1}{2}pS_{y}~.
\end{align}
The free layer is electrically floating, hence there is no net
charge current flowing through $\text{F}3|\text{N}$ interface,
$I_{c,3}=0$. The spin accumulation in the free layer
$\mu_{s}^{F}=\mu_{\uparrow}-\mu_{\downarrow}$, directed along
magnetization $\mathbf{m}_{3}$, is governed by the spin diffusion
equation,\cite{johnson-silsbee}
\begin{equation}
\frac{\partial^{2}\mu_{s}^{F}(z)}{\partial z^{2}}
=\frac{\mu_{s}^{F}(z)}{l_{sd}^{2}}
\label{eq:sdf}
\end{equation}
which satisfies the following boundary conditions. At the interface
the continuity of longitudinal spin current dictates
\begin{equation}
\sigma_{\uparrow}\left(\frac{\partial\mu_{\uparrow}}{\partial z}\right)_{z=0}
-\sigma_{\downarrow}\left(\frac{\partial\mu_{\downarrow}}{\partial z}\right)_{z=0}
=\frac{2e^{2}}{\hbar A}\mathbf{I}_{s,3}\cdot\mathbf{m}_{3}
\end{equation}
and the vanishing spin current at the end of the ferromagnet implies
\begin{equation}
\sigma_{\uparrow}\left(\frac{\partial\mu_{\uparrow}}{\partial z}\right)_{z=d}
-\sigma_{\downarrow}\left(\frac{\partial\mu_{\downarrow}}{\partial z}\right)_{z=d}
=0~.
\end{equation}
The solution of eq. (\ref{eq:sdf}) reads
\begin{equation}
\mu_{s}^{F}(z)=\frac{\zeta_{3}\cosh(\frac{z-d}{l_{sd}})\mathbf{m}_{3}\cdot\mathbf{S}}
{\left[\zeta_{3}+\tilde{\sigma}\tanh(\frac{d}{l_{sd}})\right]\cosh(\frac{d}{l_{sd}})}
\end{equation}
where $\zeta_{3}=g_{3}(1-p_{3}^{2})/8\pi$ characterizes the contact
$\text{F3}|\text{N}$ and $\tilde{\sigma}=\hbar
A\sigma_{\uparrow}\sigma_{\downarrow}/(e^{2}l_{sd}(\sigma_{\uparrow}+\sigma_{\downarrow}))$
describes the bulk properties of the free layer with arbitrary
$\mathbf{m}_{3}$. The conservation of spin currents in eq.
(\ref{eq:conserv}) generates three linear equations that determine
the spin accumulation $\mathbf{S}$ in the normal metal as
\begin{equation}
\mathbf{S}=\hat{\mathbf{\Pi}}(g,g_{3})\left(8\pi\mathbf{I}_{s}^{(p)}+\mathbf{W}_{b}\right)
\label{eq:spinaccu}
\end{equation}
where the vector $\mathbf{W}_{b}=(0,2hpI_{0}/e,0)$ is the
contribution from the bias current and the elements of the symmetric
matrix $\hat{\mathbf{\Pi}}(g,g_{3})$ is listed in the Appendix. Equation
(\ref{eq:torquegeneral}) determines the spin transfer torque acting
on the free layer magnetization, which can be arranged as
\begin{equation}
\mathbf{L}=\frac{\eta_{3}g_{3}}{8\pi}\hat{\mathbf{\Gamma}}(g,g_{3})
\left(8\pi\mathbf{I}_{s}^{(p)}+\mathbf{W}_{b}\right)
\label{eq:spintorque}
\end{equation}
and the components of the matrix $\hat{\mathbf{\Gamma}}(g,g_{3})$
are listed in the Appendix.

\section{\label{sec:thermal}Thermal Stability}
The spin transfer torque rotates the magnetization out of the
equilibrium hence increasing the magnetostatic energy
$E_{\text{MS}}$. The initial magnetization is stable against thermal
fluctuations when
\begin{equation}
\Delta E_{\text{MS}}> k_{B}T,
\end{equation}
where $k_{B}$ is the Boltzmann constant and $T$ the temperature. For
an elliptic permalloy film, disregarding any residual crystalline
anisotropy, the effective field due to the shape anisotropy can be
written as
\begin{equation}
{\bf H}_{e}=-\mu_{0}M_{s}(N_{x}m_{x},N_{y}m_{y},N_{z}m_{z})~,
\label{eq:eff-field}
\end{equation}
introducing the saturation magnetization $M_{s}$ and demagnetizing
factors $N_{x}$, $N_{y}$ and $N_{z}$~\cite{osborn}.
When the magnetization is slightly out of plane, the difference between
the magnetostatic energy for magnetizations along the hard-axis and
easy-axis reads $\Delta E_{\text{MS}}=\mu_{0}VM_{s}^{2}(N_{y}-N_{x})/2$.
For a very flat ellipsoid (the thickness is much smaller than the lateral dimensions)
and slight ellipticity (large aspect ratio $\xi\approx 1$),
we can expand the demagnetizing factors at $\xi=1$ such that
\begin{equation}
N_{y}-N_{x}=\frac{\pi d}{4a}\frac{(\xi^{2}+4\xi+1)(1-\xi)}{\xi(\xi+1)^{2}}~,
\label{eq:diffdemag}
\end{equation}
where $a$, $b$ and $d$ are the lengths of easy-axis, hard axis and
the thickness of the permalloy film. The aspect ratio is defined as
$\xi=b/a$. The requirement $\Delta E_{\text{MS}}>k_{B}T$ gives
\begin{equation}
\frac{(\xi^{2}+4\xi+1)(1-\xi)}{(\xi+1)^{2}}>
\frac{8k_{B}T}{\mu_{0}M_{s}^{2}\pi^{2}ad^{2}}~.
\label{eq:therstability}
\end{equation}
The saturation magnetization of permalloy is $M_{s}=8\times
10^{5}~\text{A}~\text{m}^{-1}$. For thickness $d=5~\text{nm}$ and
easy axis $a=200~\text{nm}$\cite{vanwees}, the right hand side of
eq. (\ref{eq:therstability}) is at room temperature approximately
$\epsilon=8.36\times 10^{-3}$ and therefore stability requires that
\begin{equation}
\xi\leq 1-\frac{2}{3}\epsilon~,
\end{equation}
which suggests that even for almost circular permalloy discs, e.g.,
$\xi=0.9$, thermal fluctuations around the equilibrium configuration
are small.

\section{\label{sec:dyna}Magnetization Dynamics}
Here we focus on the free layer magnetization dynamics in the
macrospin model. The Landau-Lifshitz-Gilbert (LLG) equation modified
by the spin transfer torque [eq. (\ref{eq:spintorque})] reads
\begin{equation}
\frac{1}{\gamma}\frac{d\mathbf{m}}{dt}=-\mathbf{m}\times\mathbf{H}_{e}
+\frac{\alpha_{0}}{\gamma}\mathbf{m}\times\frac{d\mathbf{m}}{dt}
+\frac{1}{VM_{s}}\mathbf{L}~.
\end{equation}
Included in the torque term, i.e., Eq. (\ref{eq:spintorque}), an
expression
\begin{equation}
\overleftrightarrow{\alpha}^{\prime}\equiv
\frac{\gamma\hbar \eta_{3}^{2}g_{3}^{2}}{8\pi V M_{s}}\hat{\mathbf{\Gamma}}(g,g_{3})
\end{equation}
appears as an enhancement of the Gilbert damping,\cite{yaroslavprl}
which depends on the direction of the magnetization and shows tensor
property of the pumping induced damping enhancement.\cite{xuhui}
When the conductance at the $\text{F3}|\text{N}$ contact is much
larger than the source-drain contacts and the spin flip in the
normal metal is negligible, i.e., $g_{3}\gg g$ and $g_{3}\gg
g_{sf}$, the tensor $\overleftrightarrow{\alpha}^{\prime}$ converges
to\cite{xuhui}
\begin{equation}
\frac{\gamma \hbar \text{Re}g_{3}^{\uparrow\downarrow}}{4\pi VM_{s}}
\left( %
\begin{array}{ccc}
1-m_{x}^{2} & -m_{x}m_{y} & -m_{x}m_{z}\\
-m_{x}m_{y} & 1-m_{y}^{2} & -m_{y}m_{z}\\
-m_{x}m_{z} & -m_{y}m_{z} & 1-m_{z}^{2}
\end{array}
\right)~, %
\label{eq:en-damping}
\end{equation}
which in the LLG equation reduces to a diagonal matrix
\begin{equation}
\overleftrightarrow{\alpha}^{\prime}
=\frac{\gamma \hbar \text{Re}g_{3}^{\uparrow\downarrow}}{4\pi VM_{s}}
\hat{\mathbf{1}}
\label{eq:diagonaldamp}
\end{equation}
and the coefficient in front of the matrix is exactly the value
derived for the single $\text{F}|\text{N}$
junction\cite{yaroslavprl}. In the same limit, the bias-driven term
of the torque reads,
\begin{equation}
\frac{1}{VM_{s}}\mathbf{L}_{b}=\frac{\hbar p I_{0}}{2VM_{s}|e|}
\left( %
\begin{array}{c}
-m_{x}m_{y}\\
1-m_{y}^{2}\\
-m_{y}m_{z}
\end{array}
\right)~. %
\end{equation}
In the following discussions,
we denote the enhanced Gilbert damping parameter
as $\alpha=\alpha_{0}+\alpha^{\prime}$, where $\alpha^{\prime}$
is the diagonal entry in eq.(\ref{eq:diagonaldamp}).
The LLG equation then reads
\begin{equation}
\frac{1}{\gamma}\frac{d\mathbf{m}}{dt}=-\mathbf{m}\times\mathbf{H}_{e}
+\frac{\alpha}{\gamma}\mathbf{m}\times\frac{d\mathbf{m}}{dt}
+\frac{1}{VM_{s}}\mathbf{L}_{b}~.
\label{eq:llg-full}
\end{equation}
For ultrathin permalloy films, without external field and
crystalline anisotropy, the magnetization is confined in the plane
by the shape anisotropy field given by eq.~(\ref{eq:eff-field}).
Equation (\ref{eq:llg-full}) is a nonlinear differential equation that
can be reformulated as
\begin{equation}
\frac{d\mathbf{m}}{dt}=\mathbf{f}(\mathbf{m},I_{0})
\end{equation}
where $\mathbf{f}(\mathbf{m},I_{0})$ is a vector function of
magnetization $\mathbf{m}$ and bias current $I_{0}$. According to
the theory of differential equations\cite{perko}, we find two
``equilibrium points" at which $d\mathbf{m}/dt$ vanishes
$\tilde{\mathbf{m}}_{1}=(1,0,\hbar
pI_{0}/[2e\mu_{0}VM_{s}^{2}(N_{z}-N_{x})])$ and
$\tilde{\mathbf{m}}_{2}=(0,1,0)$. Expanding eq. (\ref{eq:llg-full})
at point $\tilde{\mathbf{m}}_{2}$ and keeping only the first-oder
derivatives with respect to the magnetization, i.e.,
\begin{equation}
\frac{d\mathbf{m}}{dt}\approx
\left(\frac{\partial \mathbf{f}}{\partial \mathbf{m}}\right)_{\tilde{\mathbf{m}}_{2}}~,
\label{eq:secondcp}
\end{equation}
where $\partial\mathbf{f}/\partial\mathbf{m}$ is a matrix with
elements given by $\partial f_{i}/\partial m_{j}$. Equation
(\ref{eq:secondcp}) has non-zero solution when
\begin{equation}
\text{det}\left[\left(\frac{\partial \mathbf{f}}{\partial \mathbf{m}}
\right)_{\tilde{\mathbf{m}}_{2}}\right]=0~.
\end{equation}
This determines the critical current that is
necessary to obtain the maximum in-plane rotation, i.e., $\pi/2$:
\begin{equation}
I_{c}=\frac{2e\mu_{0}VM_{s}^{2}\sqrt{(N_{z}-N_{y})(N_{y}-N_{x})}}{\hbar p}
\label{eq:cricurrent}
\end{equation}
The LLG equation augmented by the spin transfer torque
for the present configuration suggests a two-state behavior
of the magnetization: Below the critical current $I_{c}$,
the magnetization is pushed out of the initial position (easy axis),
then undergoing damped precessions and finally stops along the easy axis
but with a small $z$-component, i.e., the equilibrium given by $\tilde{\mathbf{m}}_{1}$.
At that position, the demagnetizing field is balanced by the spin torque.
Above the critical current, the magnetization precesses out of the
easy axis and rotates to the hard axis without any precession.

\begin{figure}
\begin{center}
\includegraphics[scale=0.45]{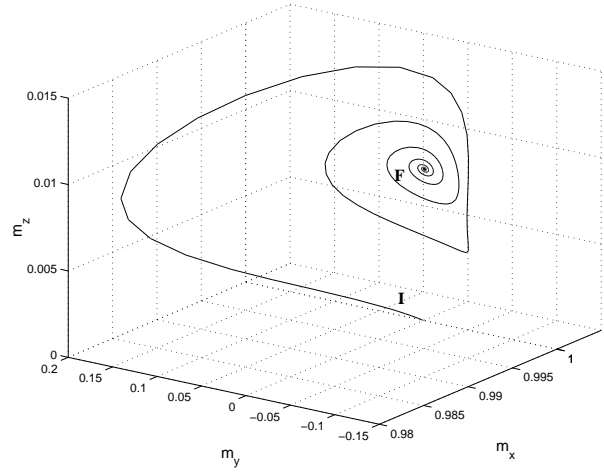}
\end{center}
\caption{\label{fig:xyz-30ma} The trajectory of the magnetization
with the bias current $I_{0}=30~\text{mA}$, which is below the critical
current. The magnetization initially aligned along easy axis ($x$-axis)
and after the damped oscillation it stops along the easy axis with
small out-of-plane component.}
\end{figure}

\begin{figure}
\begin{center}
\includegraphics[scale=0.45]{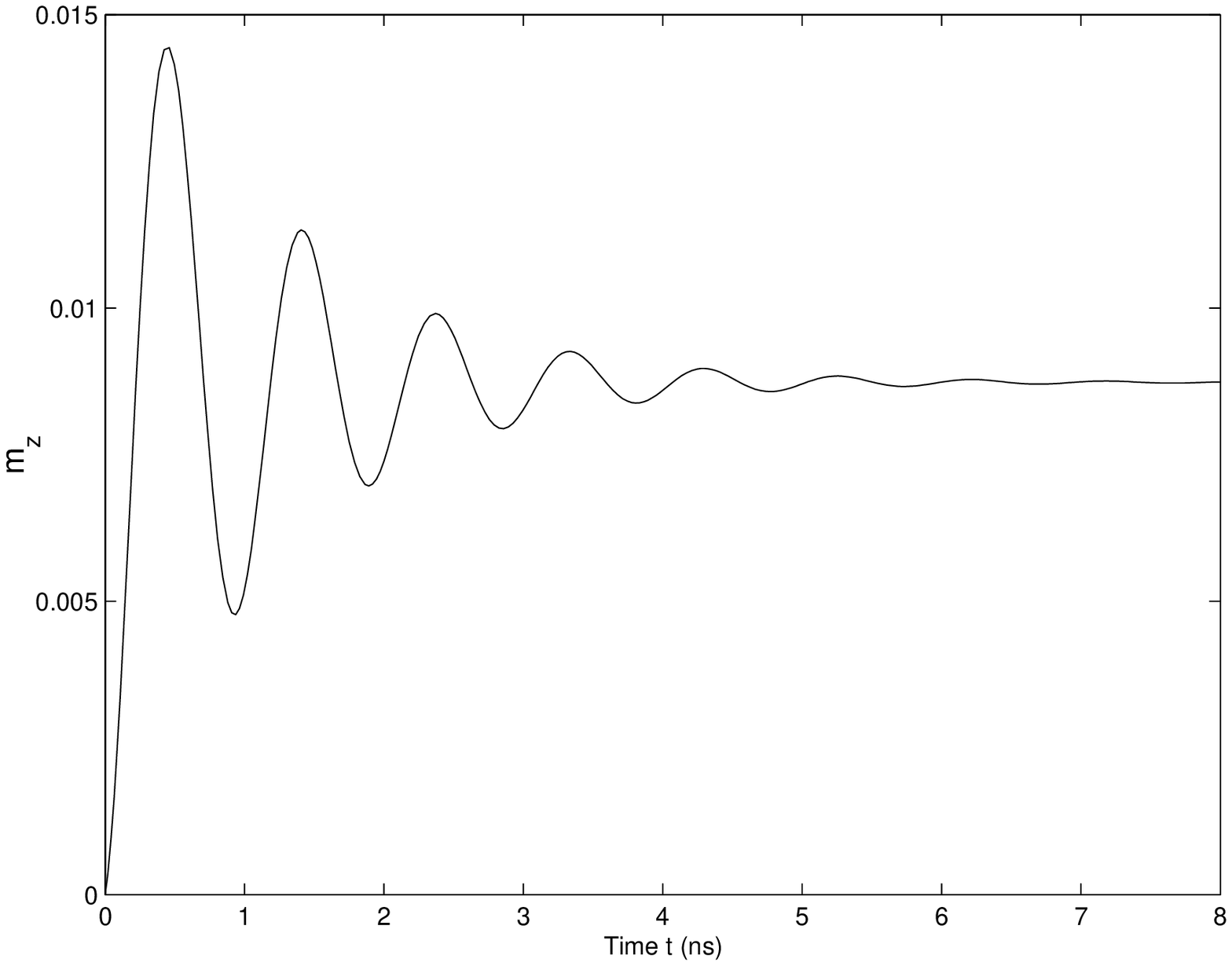}
\end{center}
\caption{\label{fig:z-30ma} The $z$-component of the magnetization vs
time (in ns). The bias current is $30~\text{mA}$.}
\end{figure}

\begin{figure}
\begin{center}
\includegraphics[scale=0.45]{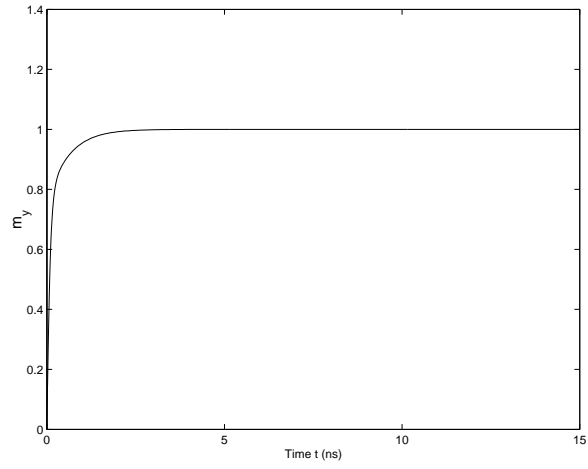}
\end{center}
\caption{\label{fig:y-160ma} The $y$-component of the magnetization vs time.
The bias current is $160~\text{mA}$, which is above the critical current.}
\end{figure}

\begin{figure}
\begin{center}
\includegraphics[scale=0.45]{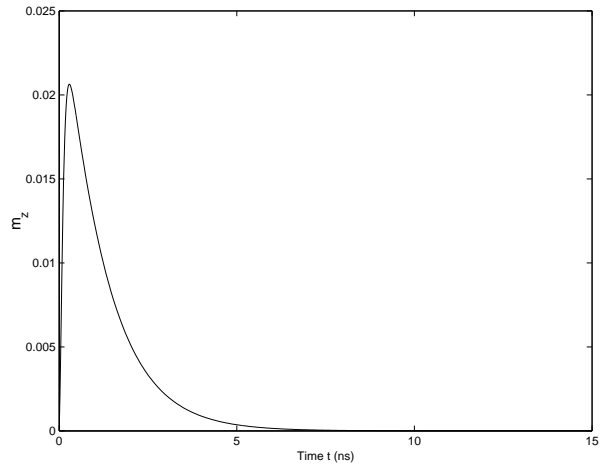}
\end{center}
\caption{\label{fig:z-160ma} The $z$-component of the magnetization vs time,
with bias current $160~\text{mA}$ that is above the critical current.}
\end{figure}

\begin{figure}
\begin{center}
\includegraphics[scale=0.45]{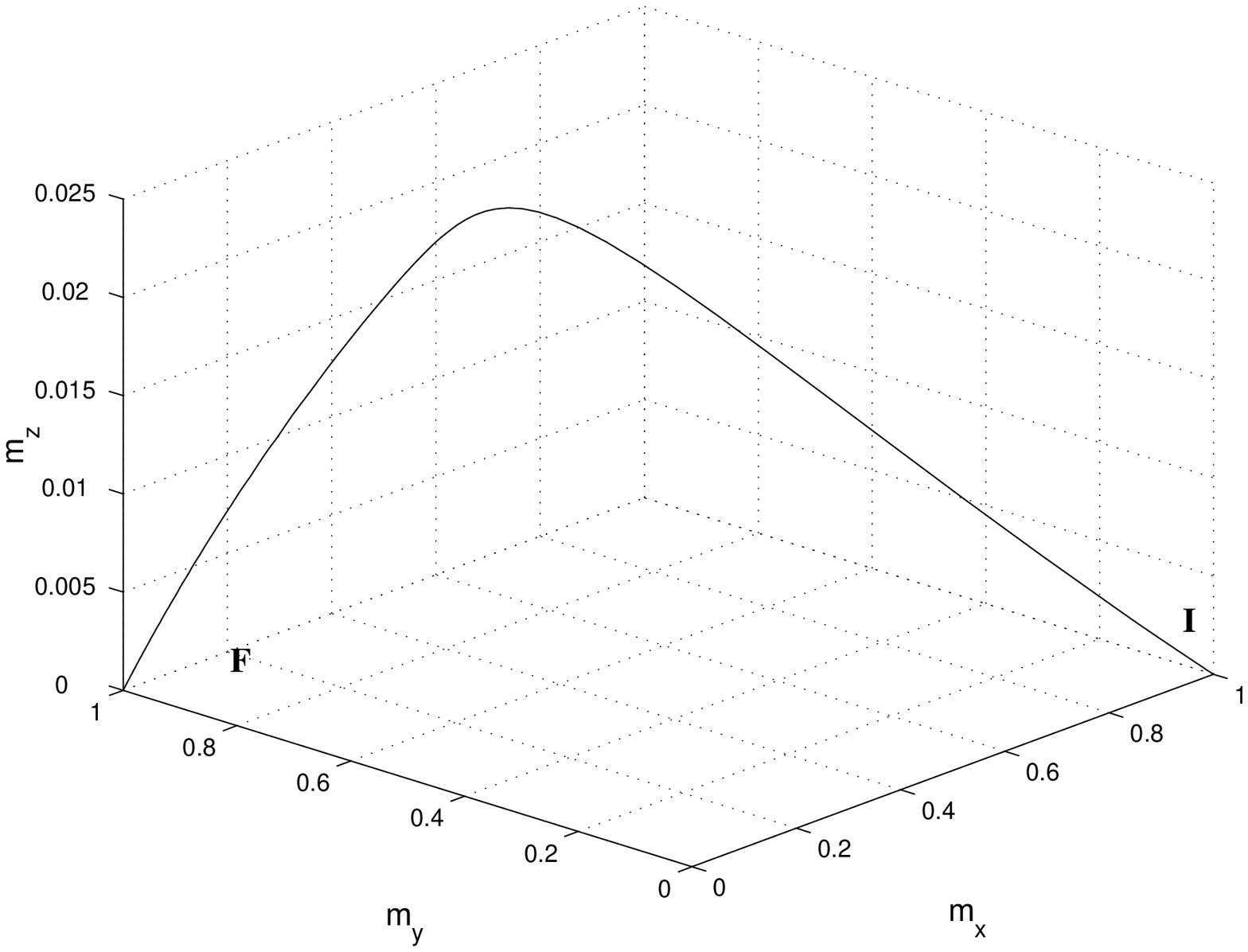}
\end{center}
\caption{\label{fig:xyz-160ma} The trajectory of the magnetization
under the bias above the critical current, $I_{0}=160~\text{mA}$.
The magnetization initially aligned along easy axis (point
\textbf{I}).}
\end{figure}

We simulate the magnetization dynamics for a polarization $p=0.4$ of
the $\text{F3}|\text{N}$ and a real part of the mixing conductance
$\text{Re}g_{3}^{\uparrow\downarrow}A^{-1}=4.1\times
10^{15}~\text{cm}^{2}$.\cite{kelly} The long semi-axis, short
semi-axis and the thickness of the permalloy island are
$a=200~\text{nm}$, $b=190~\text{nm}$, and $d=5~\text{nm}$,
respectively. The calculated demagnetizing factors are
$N_{y}=0.0224$ and $N_{x}=0.0191$.\cite{osborn} The single-spin
density of states in the normal metal is $\nu_{DOS}= 2.4\times
10^{28}~\text{eV}^{-1}\text{m}^{-3}$.\cite{vanwees}  
The bulk value of the Gilbert damping
parameter is $\alpha_{0}=0.006$ and the calculated enhancement of
Gilbert damping is $\alpha^{\prime}=0.015$.\cite{yaroslavprl}
According to eq. (\ref{eq:cricurrent}), for the above dimensions,
the critical current to achieve $\phi=\pi/2$ is
$I_{c}=139~\text{mA}$, which agrees well with the numerical results.
Below the critical current, e.g., $I_{0}=30~\text{mA}$, the
equilibrium $z$-component determined by the expression of
$\tilde{\mathbf{m}}_{1}$is $0.0087$, which also agrees with the
numerical results shown in Fig. \ref{fig:z-30ma}. The trajectory of
the magnetization when suddenly switching on the bias current
$I_{0}=30~\text{mA}$ is depicted in Fig. \ref{fig:xyz-30ma}. The
magnetization starts from the easy axis (point \textbf{I} in the
figure), undergoes a damped oscillation and finally stops at point
\textbf{F}, where the spin transfer torque induced by the spin
accumulation in the normal metal is balanced by the torque generated
by the anisotropy field. Fig. \ref{fig:xyz-160ma} shows the
trajectory of the magnetization under switching on a the bias
current $I_{0}=160~\text{mA}$, which is above the critical current.
Figures \ref{fig:y-160ma} and \ref{fig:z-160ma} are the time
dependence of the $y$ and $z$-components of the free layer
magnetization. These figures indicate that the magnetization
response to a large current is close to a step function.
A smaller size of the permalloy film requires a smaller critical
current, as indicated by eq.~(\ref{eq:cricurrent}). In the above
simulation, we did not take into account the effect of a finite RC
time for switching on the bias current. A longer rising time of the
bias current implies that it takes longer before the magnetization
reaches the steady state position. But the magnitude of the critical
current does not depend on how the bias current transient.

We finally note that with the dimensions chosen here, the bias currents
generate a significant in-plane \O rsted field that may interfere with the
spin-torque effect. It can be avoided, e.g., by spatially separating the
free layer from the current path (but within the spin-flip diffusion) \cite{kimura-condmat2005}
or by generating a neutralizing \O rsted field by a neighboring circuit (suggested by Siegmann).

The advantage of the proposed device mainly comes from the two-state
behavior separated by the critical current, which can be utilized as
the $0$ and $1$ states in current controlled memory elements. We
notice that after the magnetization being switched to the hard axis,
only small current is needed to maintain the position stable against
thermal fluctuations. Another possible application could be the
implementation of such device into spin-torque transistors
\cite{bauer-spintorque} to achieve the gain of current since the
angle of the magnetization in the above device is tunable by the
bias. The magnetization can be also used as a spin battery that is
``charged" in the high energy state (hard axis) and relaxes a spin
current into the normal metal when relaxing to the ground state
(easy axis). The induced spin accumulation then creates voltage
difference over the source and drain contacts.

\begin{figure}
\begin{center}
\includegraphics[scale=0.45]{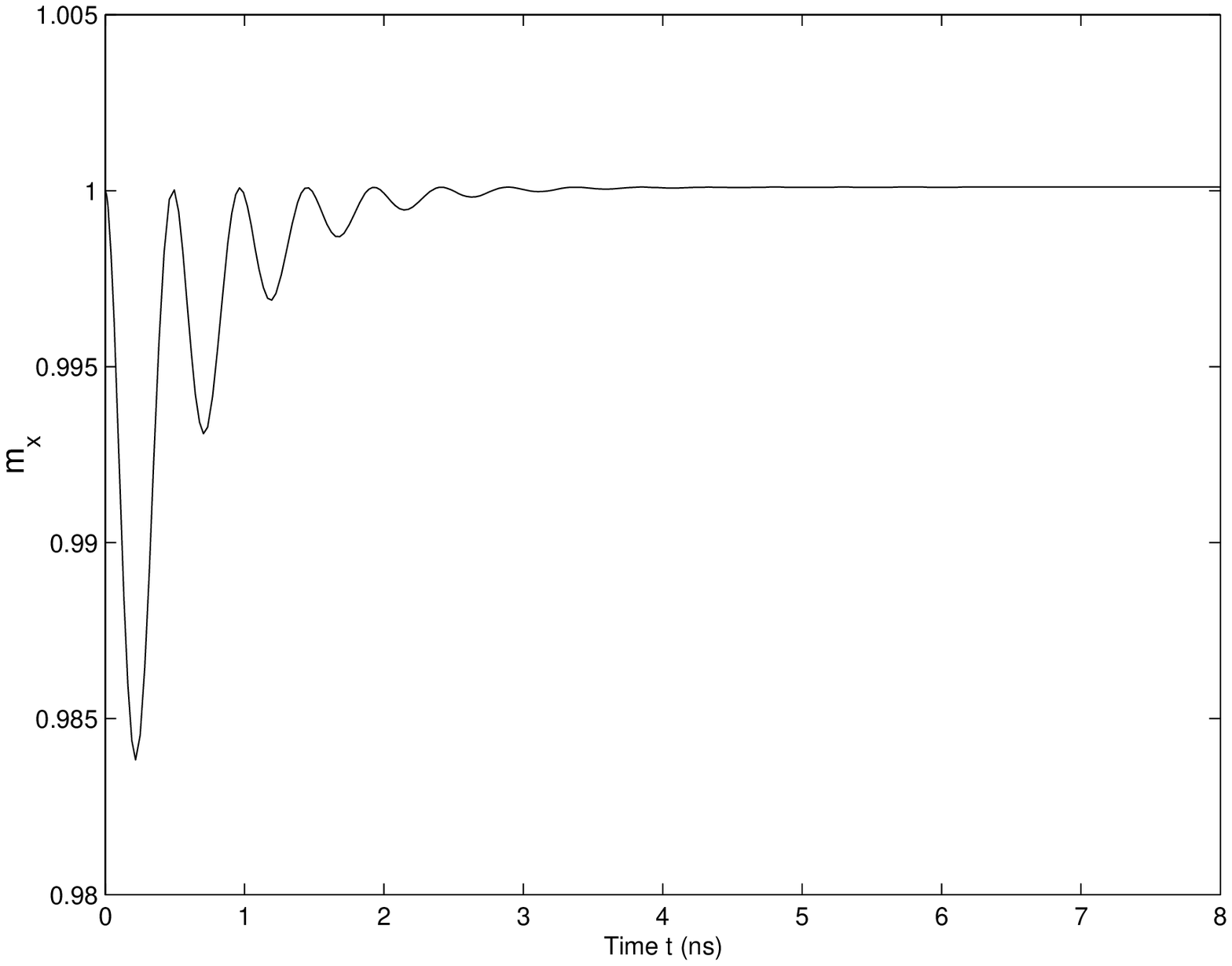}
\end{center}
\caption{\label{fig:x-30ma} The $x$-component of the magnetization vs time
(in ns). The bias current is $30~\text{mA}$.}
\end{figure}

\section{\label{sec:concl}Conclusions}
In this article, the magnetization dynamics of a spin-flip transistor
has been studies in the macrospin LLG equation combined with MECT.
We found a two-state behavior of the free layer magnetization controlled
by the current induced spin transfer torque and spin pumping. The two
regimes are separated by a critical current, below which the magnetization
undergoes a damped oscillation and stops along the easy axis with small $z$-component.
Above the critical current, the magnetization rotates to the hard ($y$-)
axis without precession. The critical current is found to depend on the size of the
free layer, the aspect ratio of the ellipsoid, and the source-drain contact polarizations.
The thermal instability analysis indicates that at room temperature the predicted effects
are visible even for very large aspect ratios.

\begin{acknowledgments}
We thank J. C. Slonczewski and A. Kovalev for discussions. X. Wang thanks
H. Saarikoski for his help on MatLab.
This work is supported by NanoNed and FOM.
\end{acknowledgments}

\bibliography{sftrans}

\begin{appendix}
\section{Spin Accumulation and Spin Transfer Torque}
The elements of the symmetric matrix $\hat{\mathbf{\Pi}}(g,g_{3})$
in eq. (\ref{eq:spinaccu}) are listed in the following
\begin{align}
\mathbf{\Pi}_{11}=&[\eta_{3}g_{3}+2g_\text{sf}+2(1-p^{2})g][2\eta g+\eta_{3}g_{3}+2g_\text{sf}-g_{3}(\eta_{3}-(1-p_{3}^{2})(1-\Delta_{3}))m_{z}^{2}]
/\mathcal{G}\\\nonumber&-(2\eta g+\eta_{3}g_{3}+2g_\text{sf})g_{3}[\eta_{3}-(1-p_{3}^{2})(1-\Delta_{3})]m_{y}^{2}/\mathcal{G}\\
\mathbf{\Pi}_{12}=&\mathbf{\Pi}_{21}=(2\eta g+\eta_{3}g_{3}+2g_\text{sf})g_{3}[\eta_{3}-(1-p_{3}^{2})(1-\Delta_{3})]m_{x}m_{y}/\mathcal{G}\\
\mathbf{\Pi}_{13}=&\mathbf{\Pi}_{31}=[\eta_{3}g_{3}+2g_\text{sf}+2(1-p^{2})g]g_{3}[\eta_{3}-(1-p_{3}^{2})(1-\Delta_{3})]m_{x}m_{z}/\mathcal{G}\\
\mathbf{\Pi}_{22}=&(2\eta g+\eta_{3}g_{3}+2g_\text{sf})
[2\eta g+\eta_{3}g_{3}+2g_\text{sf}-g_{3}(\eta_{3}-(1-p_{3}^{2})(1-\Delta_{3}))(m_{x}^{2}+m_{z}^{2})]/\mathcal{G}\\
\mathbf{\Pi}_{23}=&\mathbf{\Pi}_{32}=(2\eta g+\eta_{3}g_{3}+2g_\text{sf})g_{3}[\eta_{3}-(1-p_{3}^{2})(1-\Delta_{3})]m_{y}m_{z}/\mathcal{G}\\
\mathbf{\Pi}_{33}=&[\eta_{3}g_{3}+2g_\text{sf}+2(1-p^{2})g][2\eta g+\eta_{3}g_{3}+2g_\text{sf}-g_{3}(\eta_{3}-(1-p_{3}^{2})(1-\Delta_{3}))m_{x}^{2}]
/\mathcal{G}\\\nonumber &-(2\eta g+\eta_{3}g_{3}+2g_\text{sf})g_{3}[\eta_{3}-(1-p_{3}^{2})(1-\Delta_{3})]m_{y}^{2}/\mathcal{G}
\end{align}
where we have introduced the following notation
\begin{align}
\mathcal{G}=(2\eta g+\eta_{3}g_{3}+2g_\text{sf})&
[(\eta_{3}g_{3}+2g-2p^{2}g+2g_\text{sf})(2\eta g+2g_\text{sf}+(1-p_{3}^{2})(1-\Delta_{3})g_{3})\\\nonumber
&+2(p^{2}-1+\eta)g(\eta_{3}-(1-p_{3}^{2})(1-\Delta_{3}))g_{3}m_{y}^{2}]~,
\end{align}
and
\begin{equation}
\Delta_{3}=\frac{\zeta_{3}}{\zeta_{3}+\tilde{\sigma}\tanh(d/l_\text{sd})}~.
\end{equation}
The matrix contained in the expression of spin transfer torque eq. (\ref{eq:spintorque})
has the following components
\begin{align}
\mathbf{\Gamma}_{11}&=[\eta_{3}g_{3}+2g_\text{sf}+2(1-p^{2})g][2\eta g+\eta_{3}g_{3}+2g_\text{sf}-g_{3}(\eta_{3}-(1-p_{3}^{2})(1-\Delta_{3}))m_{z}^{2}]
/\mathcal{G}\\\nonumber
&-(2\eta g+\eta_{3}g_{3}+2g_\text{sf})
[(\eta_{3}-(1-p_{3}^{2})(1-\Delta_{3}))g_{3}m_{y}^{2}+(\eta_{3}g_{3}+2g_\text{sf}+2(1-p^{2})g)m_{x}^{2}]/\mathcal{G}\\
\mathbf{\Gamma}_{12}&=-(2\eta g+\eta_{3}g_{3}+2g_\text{sf})[2\eta g+2g_\text{sf}+(1-p_{3}^{2})(1-\Delta_{3})g_{3}]m_{x}m_{y}/\mathcal{G}\\
\mathbf{\Gamma}_{13}&=-[\eta_{3}g_{3}+2g_\text{sf}+2(1-p^{2})g][2\eta g+2g_\text{sf}+(1-p_{3}^{2})(1-\Delta_{3})g_{3}]m_{x}m_{z}/\mathcal{G}\\
\mathbf{\Gamma}_{21}&=-(2\eta g+\eta_{3}g_{3}+2g_\text{sf})[2g_\text{sf}+2(1-p^{2})g+(1-p_{3}^{2})(1-\Delta_{3})g_{3}]m_{x}m_{y}/\mathcal{G}\\
\mathbf{\Gamma}_{22}&=(2\eta g+\eta_{3}g_{3}+2g_\text{sf})[2\eta g+2g_\text{sf}+(1-p_{3}^{2})(1-\Delta_{3})g_{3}](1-m_{y}^{2})/\mathcal{G}\\
\mathbf{\Gamma}_{23}&=-(2\eta g+\eta_{3}g_{3}+2g_\text{sf})[2g_\text{sf}+2(1-p^{2})g+(1-p_{3}^{2})(1-\Delta_{3})g_{3}]m_{y}m_{z}/\mathcal{G}\\
\mathbf{\Gamma}_{31}&=-[\eta_{3}g_{3}+2g_\text{sf}+2(1-p^{2})g][2\eta g+2g_\text{sf}+(1-p_{3}^{2})(1-\Delta_{3})g_{3}]m_{x}m_{z}/\mathcal{G}\\
\mathbf{\Gamma}_{32}&=-(2\eta g+\eta_{3}g_{3}+2g_\text{sf})[2\eta g+2g_\text{sf}+(1-p_{3}^{2})(1-\Delta_{3})g_{3}]m_{y}m_{z}/\mathcal{G}\\
\mathbf{\Gamma}_{33}&=[\eta_{3}g_{3}+2g_\text{sf}+2(1-p^{2})g][2\eta g+\eta_{3}g_{3}+2g_\text{sf}-g_{3}(\eta_{3}-(1-p_{3}^{2})(1-\Delta_{3}))m_{x}^{2}]
/\mathcal{G}\\\nonumber
&-(2\eta g+\eta_{3}g_{3}+2g_\text{sf})
[(\eta_{3}-(1-p_{3}^{2})(1-\Delta_{3}))g_{3}m_{y}^{2}+(\eta_{3}g_{3}+2g_\text{sf}+2(1-p^{2})g)m_{z}^{2}]/\mathcal{G}~.
\end{align}

\end{appendix}

\end{document}